\documentclass[aps,pre,twocolumn,superscriptaddress,10pt]{revtex4-2}
\pdfoutput=1

\usepackage{amsmath}
\usepackage{bm}
\usepackage{caption}
\usepackage{graphicx}
\usepackage{microtype}
\usepackage{xcolor}

\usepackage[
    colorlinks=true,
    linkcolor=blue,
    citecolor=black,
    urlcolor=black
]{hyperref}

\usepackage{cleveref}
\usepackage{amsmath}

\makeatletter
\AfterEndEnvironment{strip}{%
  \par\@afterindentfalse\@afterheading
}
\makeatother

\NewDocumentCommand{\SI}{o}{%
    \IfNoValueTF{#1}
    {\textit{SI Appendix}}
    {(see \textit{SI Appendix}, #1)}
}

\NewDocumentCommand{\SInoref}{}{(see \textit{SI Appendix})}

\usepackage{xspace}

\NewDocumentCommand{\MM}{o}{%
  \IfNoValueTF{#1}
    {\textit{Materials and Methods}}
    {(see \textit{Materials and Methods}, #1)}%
  \xspace
}

\NewDocumentCommand{\MMnoref}{}{(see \textit{Materials and Methods})}

\begin{document}

\title{A multiscale theory based on metabolic scaling connects forest dynamics to tree-size distributions}

\newcommand{\DFA}{Laboratory of Interdisciplinary Physics, Department of Physics and Astronomy ``G. Galilei", University of Padova, Padova, Italy}
\newcommand{\TESAF}{Department TESAF, University of Padova, Padova, Italy}
\newcommand{\NBFC}{National Biodiversity Future Center, Palermo, Italy}
\newcommand{\INFN}{INFN, Sezione di Padova, via Marzolo 8, 35131 Padova, Italy}
\newcommand{\DICEA}{Department of Civil, Environmental and Architectural Engineering, University of Padova, Padova, Italy}
\newcommand{\CMCC}{CMCC Centro Euro-Mediterraneo sui Cambiamenti Climatici, Lecce, Italy}

\author{Christian Grilletta}
\affiliation{\DFA}

\author{Tommasso Anfodillo}
\affiliation{\TESAF}

\author{Gaia Pasqualotto}
\affiliation{\TESAF}
\affiliation{\NBFC}

\author{Samir Suweis}
\affiliation{\DFA}
\affiliation{\INFN}

\author{Andrea Rinaldo}
\email{andrea.rinaldo@unipd.it}
\affiliation{\DICEA}
\affiliation{\CMCC}

\author{Amos Maritan}
\email{amos.maritan@unipd.it}
\affiliation{\DFA}
\affiliation{\NBFC}
\affiliation{\INFN}

\author{Davide Bernardi}
\email{davide.bernardi@unipd.it}
\affiliation{\DFA}
\affiliation{\NBFC}
\affiliation{\INFN}

\begin{abstract}
Scaling relations linking species size, abundance, and resource availability are among the most robust empirical regularities in ecology. However, a mechanistic explanation for how these community-level laws emerge from ecological processes remains elusive.  Here, we address this gap by developing a minimal spatially explicit dynamical framework for forest communities that incorporates seed dispersal, growth limited by local light availability, local competition, and global resource constraints grounded in metabolic scaling principles.  By deriving an analytical solution for the tree-size distribution, we show that its stationary state exhibits two distinct power-law regimes whose exponents are controlled by the relative strength of resource and spatial competition. The crossover between these regimes is set by the interplay between seed injection and local resource availability, establishing an explicit link between the scaling exponent of the size distribution and forest condition. Finally, we show that boundary disturbances can break the ecological balance between competing species and induce effects that propagate deeply into the forest bulk, far beyond the single-plant dispersal range. Together, these results provide a unifying dynamical perspective on forest scaling laws with potential applications to a broad range of biological communities.
\end{abstract}

\maketitle

\section*{Introduction}
Ecological systems exhibit a remarkable variety of macroscopic regularities despite the complexity of the underlying biological interactions \cite{Levin1992}. Understanding how such large-scale patterns emerge from the dynamics of individual organisms is a central challenge in ecology. Among the descriptors of ecological communities, organism size occupies a particularly important role because it influences metabolism, growth, reproduction, resource use, and competitive ability  \cite{Kleiber1961energetics,damuth2001scaling,Marquet_2005}. As a consequence, size distributions provide a natural link between individual-level processes and ecosystem-level structure  \cite{Damuth1981,banavar1999size,enquist2001invariant,Marquet_2002}, and explaining the origin of these distributions and their associated scaling laws has therefore been a longstanding objective in ecological theory.

A major advance in this direction came from allometric scaling theory and the Metabolic Theory of Ecology (MTE), which revealed the existence of systematic relationships linking organism size, metabolic rates, growth, and demographic processes \cite{west1997general,enquist1998allometric,west2001general,enquist2002global,brown2004toward,Marquet2004}. These ideas have provided a unified framework for understanding how biological constraints operating at the level of individual organisms can generate regularities across populations and communities.
In forest ecosystems, in particular, metabolic and allometric arguments have successfully explained broad patterns of biomass allocation, resource use, and forest structure across a wide range of species and environments \cite{purves2008predicting,strigul2008scaling,mori2010mixed,banavar2014form,simini2010self,volkov2022seeing}. Consistent with these ideas, empirical tree-size distributions across forest ecosystems display approximate power-law scaling \cite{enquist1998allometric,Enquist1999,simini2010self}, which can be interpreted as a signature of self-similarity and suggests that common patterns in complex and seemingly distinct ecological systems may emerge from simple underlying constraints, such as those proposed by MTE.

An important open question is how these large-scale patterns emerge from the interaction of demographic processes operating within ecological communities. Ecosystems are inherently multiscale systems in which individual growth, competition, mortality, and recruitment continuously reshape community structure \cite{Levin1992}. Although observed size distributions constrain the class of admissible ecological theories, they do not uniquely identify the ecological processes that generate them. Bridging this gap therefore requires mechanistic dynamical models that connect individual-level biological processes to emergent community structure \cite{Xiao_JOD_2016}. 
Size-structured population models provide a natural mathematical framework for this purpose and have played a prominent role in forest ecology and population dynamics \cite{McKendrick_1925,Levin_1974,Takada1986,keyfitz1997mckendrick,kohyama1991simulating,kohyama1992density,kohyama1993size,kohyama2003tree,murray2003mathematical,muller2006comparing,o2009integrative}. At the same time, combining allometric growth, nonlinear competition, and ecological interactions within analytically tractable size-structured models remains a significant challenge.

Spatial processes introduce an additional layer of complexity. While trees are sessile organisms, recruitment occurs through seed dispersal, generating interactions that extend across spatial scales. In forest ecosystems, dispersal therefore provides the primary mechanism linking local populations across space. Understanding how nonlocal recruitment interacts with growth and competition requires frameworks that explicitly incorporate spatial interactions while remaining analytically tractable. Recent work has shown how the interplay between growth, competition, and spatial organization can produce emergent ecological patterns \cite{lee2021growth}. More broadly, spatial structure has been shown to influence community composition, ecosystem resilience, and long-term stability, while spatial patterns themselves can provide information about ecosystem condition \cite{Nathan2000,Wiegand2021,Kalyuzhny2023,Bernardi2026}.

Here we develop a spatially explicit, size-structured model for forest dynamics derived from metabolic scaling principles. Starting from an energy-balance description of individual growth, we construct a non-linear integro-differential equation governing the density of trees as a function of size, space, and time. The model couples ontogenetic growth, local effects such as light shading and density-dependent competition for space and resources, and nonlocal recruitment mediated by seed dispersal. Despite the complexity of these interacting processes, the framework remains analytically tractable and admits exact and asymptotic solutions in several biologically relevant regimes.
This allows us to show that the stationary tree-size distribution naturally exhibits a crossover between distinct scaling regimes associated with resource limitation and spatial exclusion, thereby linking observable forest structure to the dominant ecological mechanisms shaping the dynamics. We further show that spatial boundaries induce strong deviations from the homogeneous state by selectively suppressing the abundance of smaller individuals, and that in multi-species systems, these effects can generate long-term competitive imbalances between species characterized by different dispersal strategies.

\section*{Results}
\begin{figure*}[t!]
\centering
\includegraphics[width=0.9\textwidth]{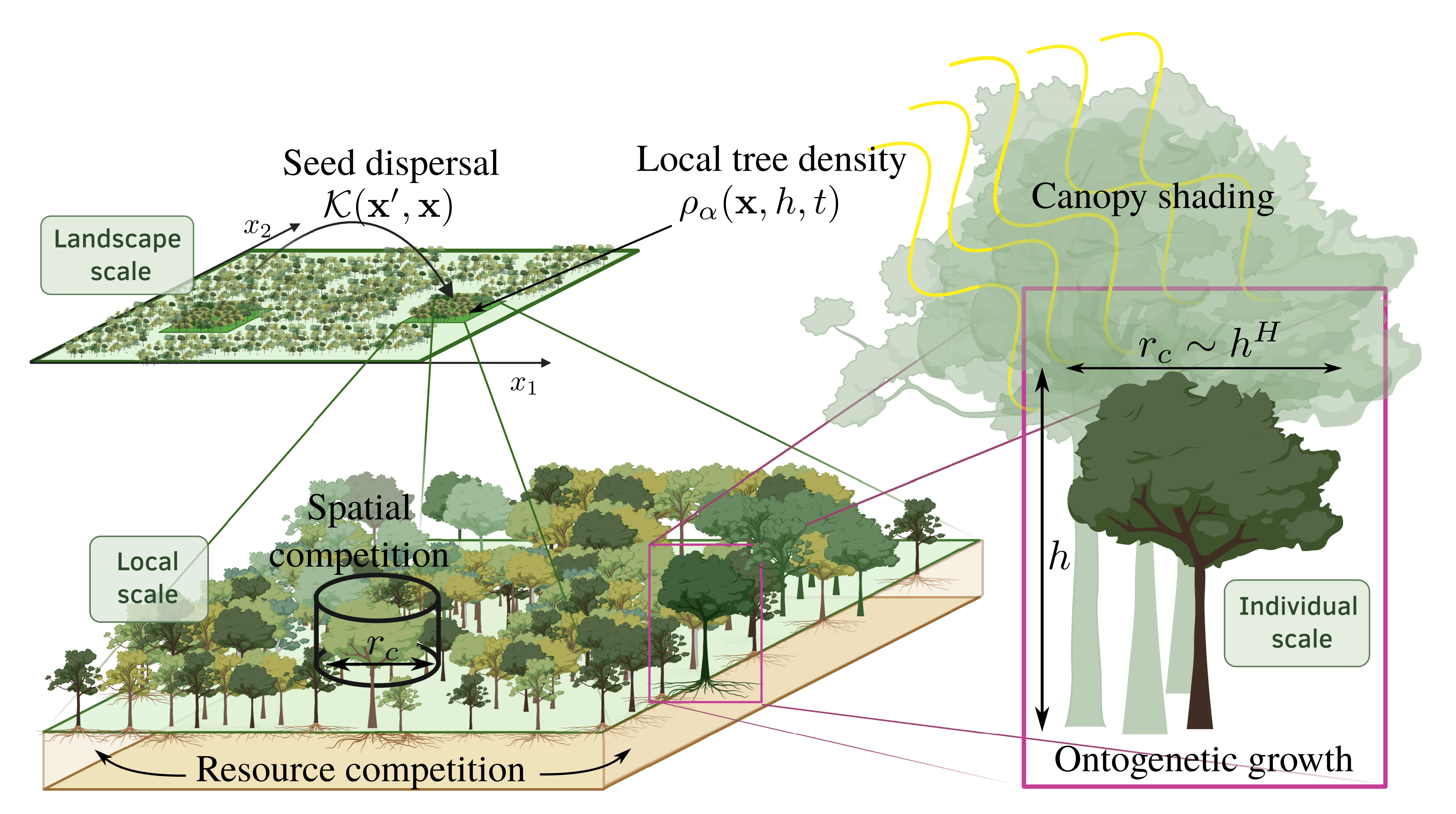}
\caption{\textbf{Schematic representation of the multiscale forest dynamics model.} 
The model describes the dynamics of the local tree density $\rho_\alpha(\mathbf{x}, h ,t)$, which represents the density per unit area of trees of species $\alpha$, size $h$, at position $\mathbf{x}=(x_1, x_2)$ and time $t$. The three panels illustrate the individual, local, and landscape scales at which the underlying processes act. The dynamics follows a McKendrick–von Foerster equation combining: (i) ontogenetic growth, derived from metabolic scaling and modulated by canopy shading through the fraction of available light; (ii) density-dependent mortality arising from competition for canopy space among similarly sized trees and from competition for shared resources among neighboring trees of all sizes; and (iii) nonlocal recruitment through seed dispersal, represented by a boundary injection term accounting for the production, dispersal, and establishment of new saplings.}
\label{fig:model}
\end{figure*}

\subsection*{A forest growth model based on metabolic scaling}
The central quantity in our model, represented schematically in \cref{fig:model}, is the local density of trees per unit area, denoted by $\rho_{\alpha}(\mathbf{x},h,t)$. Then, $\rho_{\alpha}(\mathbf{x},h,t)dh A$ represents the number of individuals of species $\alpha$ within the reference area $A$ centered at spatial position $\mathbf{x}$ at time $t$ whose height lies in the interval $(h,h+dh)$. \
We assume that recruitment occurs at a fixed height $h=h_0>0$, corresponding to a seed that has successfully survived and grown into a seedling. Individuals below $h_0$ are therefore not tracked explicitly, and $\rho_\alpha(\mathbf{x},h,t)$ is defined only for $h\geq h_0$.
As individuals grow, they are transported across height classes with growth velocity $\tilde g_{\alpha}(h,t)$, while mortality removes individuals at rate $d_{\alpha}(\mathbf{x},h,t)$. Hence, the temporal evolution of this density takes the form of a McKendrick–von Foerster equation  \cite{McKendrick_1925,keyfitz1997mckendrick,murray2003mathematical}, a continuity equation in height space:
\begin{equation}
\begin{split}
    \partial_t \rho_{\alpha}(\mathbf{x},h,t) =& - \partial_{h}\left[ \tilde{g}_{\alpha}(\mathbf{x},h,t) \rho_{\alpha}(\mathbf{x},h,t) \right]\\ &- \mu_{\alpha}(\mathbf{x},h,t) \rho_{\alpha}(\mathbf{x},h,t).
\end{split}
\label{eq:model-main}
\end{equation}
The first term, $\tilde{g}_{\alpha}(\mathbf{x},h,t)$, describes the flux of individuals through height space due to ontogenetic growth, whereas the second, $\mu_{\alpha}(\mathbf{x},h,t)$, accounts for mortality induced by competition for space and resources. Recruitment and dispersal enter through the boundary condition at small heights and through the spatial dependence of the density field, as explained below.

We first consider the growth of an isolated tree, neglecting competition and external limitations. In trees, only a fraction of the total biomass, termed sapwood, is metabolically active.
Following standard metabolic arguments, we assume that the metabolically active biomass $M$ obeys the energy balance equation \cite{ west1997general, enquist1998allometric, enquist2002global, brown2004toward, west2001general, mori2010mixed} \MMnoref.
We further assume, consistently with previous studies, that the metabolic rate is proportional to the total leaf area and therefore to the evapotranspiration rate \cite{mori2010mixed, simini2010self, banavar2014form, volkov2022seeing}. 
Because leaf area scales with crown volume, and the crown radius scales with tree height as $r_c \sim h^H$ ($0 < H \leq 1$ is the Hurst exponent \cite{simini2010self, Comita2007}), both the metabolic rate and the total sapwood mass can be related to tree height through allometric scaling relations \MMnoref. Substituting such scaling relations into the energy balance equation yields the following effective growth law for tree height:
\begin{equation}
\frac{dh_{\alpha}(t)}{d t}=g_\alpha(h)=g_0^{\alpha}\left(1-\frac{h}{h_u^{\alpha}}\right),
\label{eq:single-tree-growth}
\end{equation}
where $h^\alpha_u$ is the characteristic height at which maintenance costs balance metabolic production. This height depends on resource availability, and in the absence of competition and external disturbances it represents the maximum attainable height for the species \MMnoref. 
We now account for the reduction in light availability caused by taller individuals. Only a fraction of the incident radiation reaches smaller individuals, reducing their growth rate.  We therefore model the effective growth rate as 
\begin{equation}
\tilde{g}(\mathbf{x},h,t)=I_{\alpha}(\mathbf{x},h,t) g_{\alpha}(h),
\end{equation}
where $I_{\alpha}(\mathbf{x},h,t)$ denotes the fraction of light at height $h$ that can be exploited by species $\alpha$. Assuming that light attenuation follows the Beer-Lambert law for plant canopies \cite{monsi1953lichtfaktor,MONSI2005}, the available light decreases exponentially with the cumulative vegetation crossed by the incoming radiation:
\begin{equation}
I_{\alpha}(\mathbf{x},h,t)=I_0\exp\left\{-\gamma_{\alpha}
\int_h^{\infty}\rho(\mathbf{x},h',t)\,h'^{2H}\,dh'
\right\},
\end{equation}
where $\rho(\mathbf{x},h,t) = \sum_\alpha \rho_\alpha(\mathbf{x},h,t)$ is the total tree density (summed across species) and $\gamma_{\alpha}$ is a coefficient describing how species $\alpha$ reacts to the decreased light availability (we remark that $I_{\alpha}(\mathbf{x},h,t)$ is not the total available light, but also integrates the physiological effect of light intensity reduction on the growth rate). Unlike the resource competition term introduced below, shading acts directly on the growth rate by reducing the energy available for biomass production.

We now turn to the second term in \cref{eq:model-main}, which models density-dependent mortality arising from competition for canopy space and finite shared resources. As trees grow, both their metabolic demand and canopy area increase, so that forest dynamics become progressively constrained by the availability of light, canopy area, and other limiting resources.

We model the mortality term as the sum of two contributions, a space-limitation term and a resource-limitation term, $\mu_{\alpha}(\mathbf{x},h,t) = d_{\alpha}^S(\mathbf{x},h,t) + d_{\alpha}^R(\mathbf{x},h,t)$, where $d_\alpha^S$ accounts for mortality induced by space limitation and $d_\alpha^R$ for mortality induced by resource limitation. The space-competition contribution is defined as
\begin{equation}
d_{\alpha}^S(\mathbf{x},h,t) = d_{\alpha} I_\alpha(\mathbf{x}, h, t) h^{2H} \rho(\mathbf{x},h,t),
\label{eq:death-S}
\end{equation}
where the functional form of \cref{eq:death-S} assumes mortality induced by space competition to be proportional to the local density of similarly sized individuals within this interaction range and to the available light intensity. More details are provided in the \MM{}.
The second term models mortality induced by resource limitation, defined as
\begin{equation}
d_{\alpha}^R(\mathbf{x},h,t) = \frac{\tilde{g}_{\alpha}(h)}{h} \frac{R\left[ \rho(\mathbf{x},t) \right]}{R_a(\mathbf{x})},
\label{eq:death-R}
\end{equation}
where $R_a(\mathbf{x})$ denotes the locally available resource flux, and
\begin{equation}
R\left[ \rho(\mathbf{x},t) \right]= \int_{h_0}^{\infty}\! dh\, \rho(\mathbf{x},h,t) h^{1+2H},
\label{eq:resources}
\end{equation}
represents the total resource consumption rate of the local forest community.
The quantity $R[\rho]$ is proportional to the total metabolic demand of the community, obtained by integrating the metabolic scaling $B\sim h^{1+2H}$ over all individuals at position $\mathbf{x}$. The ratio
\begin{equation}\label{eq:a}
a(\mathbf{x},t) = \frac{R[\rho(\mathbf{x},t)]}{R_a(\mathbf{x})}
\end{equation}
therefore measures the local level of resource saturation. The form of \cref{eq:death-R} expresses the idea that mortality due to resource competition should increase both with the metabolic demand of an individual and with the degree of collective resource depletion.
In the absence of additional effects, the resulting dynamics drives the system toward a stationary state in which the total resource consumption equilibrates around the available resource flux $R_a$ \MMnoref. Although we treat resources here as an effective collective variable, all species drain resources from the shared resource pool, thereby generating both intra- and interspecific competition.

Finally, we model the recruitment of new individuals through a boundary condition at the sampling height $h_0$: 
\begin{align}\label{eq:injection-rate}
\!\!\rho_{\alpha}(\mathbf{x},h_0,\!t)\!=\!\int_{h_b}^{\infty}\!\!\!dh\!\!\int\!\!d\mathbf{x}' \mathcal K_\alpha(\mathbf{x},\mathbf{x}') h^{1+2H} \rho_{\alpha}(\mathbf{x}'\!,h,t).
\end{align}
Here, $\mathcal{K}(\mathbf{x},\mathbf{x}')$ 
denotes the dispersal kernel, which quantifies the rate at which an individual located at $\mathbf{x}'$ contributes viable offspring to position $\mathbf{x}$.
The kernel therefore effectively incorporates both seed dispersal and germination processes, including the effects of biotic and abiotic dispersal vectors and local environmental conditions. In principle, the kernel could depend explicitly on both the time of seed production and the time of recruitment. For simplicity, here we assume a stationary kernel and consider $\mathcal{K}(\mathbf{x},\mathbf{x}')$ as an integrated effect over all past times. This approximation is further justified by the fact that seed dispersal is expected to occur on time-scales that are fast compared to those of tree birth and death dynamics.
\Cref{eq:injection-rate} expresses the density of newly established saplings as a non-local contribution from reproductively active individuals across the forest. The contribution of each individual is weighted both by the dispersal kernel and by the factor $h^{1+2H}$, which reflects the scaling of reproductive investment with metabolic rate.
The integral over tree height starts at $h_b$, representing the minimum height at which individuals become reproductively mature. For simplicity, and in order to reduce the number of free parameters, in the following we set $h_b=h_0$, while the more general case $h_b\geq h_0$ is discussed in the \SI.
Within our framework, spatial effects enter the dynamics exclusively through the recruitment process. This reflects the fact that trees are sessile organisms, while seed dispersal can occur over finite distances through multiple biotic and abiotic transport mechanisms. Such non-local interactions are explicitly encoded in the dispersal kernel $\mathcal K(\mathbf{x}\,,\mathbf{x}')$. 

Combining all contributions (see \MM for derivation), the complete dynamical equation governing the time evolution of the population density is:
\begin{widetext}
\begin{equation}
\partial_t \rho_{\alpha}\!(\mathbf{x},h,t) \!= 
-I_\alpha(\mathbf{x},h,t)
\left\{\partial_h\left[g_\alpha(h) \rho_{\alpha}(\mathbf{x},h,t)\right]
\!+\!\left[(\gamma_\alpha g_\alpha(h) + d_\alpha)h^{2H}\rho(\mathbf{x},h,t) 
\!+\!\frac{g_\alpha(h)}{h}\frac{R[\rho(\mathbf{x},t)]}{R_a(\mathbf{x})}\right]\! \rho_{\alpha}(\mathbf{x},h,t)\right\},
\label{eq:complete-growth-equation}
\end{equation}
\end{widetext}
which, complemented by the injection term \cref{eq:injection-rate}, defines our model in the general case. 
We conclude this section by noting that the classical McKendrick-von Foerster equation is linear when demographic rates and recruitment are prescribed independently of the evolving population. In this setting, long-time behavior is determined by the spectral properties of the associated linear operator, so that stationary populations arise only under specific parameter balances  \cite{murray2003mathematical}. In practice, this is often achieved either by calibrating demographic parameters to satisfy the equilibrium condition or by deriving demographic rates under the assumption of a prescribed stationary distribution \cite{lee2021growth}. In contrast, our formulation couples growth, mortality, and recruitment self-consistently through ecological interactions, leading to a nonlinear nonlocal equation in which the stationary state emerges as part of the dynamics.
To systematically investigate the dynamical behavior of the model and its connection to the underlying biological mechanisms, we now proceed by first introducing several simplifying assumptions, which we will then gradually relax.

\subsection*{Exact solution of the spatially-homogeneous bulk solution}
To obtain analytical insight into the full model, we first consider the spatially homogeneous bulk limit.
First, we assume that growth and resource-consumption parameters, along with the shadow coefficient, are identical across species, in the spirit of the neutral assumption of community ecology \cite{Hubbell2001}, which successfully captures several large-scale patterns in tropical forests. Under this assumption, the multispecies dynamics can be written exactly in terms of the total density $\rho(\mathbf{x},h,t) = \sum_{\alpha}\rho_{\alpha}(\mathbf{x},h,t)$.
This approximation allows us to isolate the effects of size structure and spatial interactions before reintroducing species-specific differences below.
Second, we assume that all environmental parameters are spatially homogeneous and that the system itself is translationally invariant. The dispersal kernel, $\mathcal{K}(\mathbf{x} -\mathbf{x}')$ encodes the characteristic spatial scale $\xi$ over which seeds are dispersed.
In the limit in which the characteristic system size is much smaller than the dispersal length scale $\xi$, the dispersal kernel is effectively constant across the domain, so recruitment becomes independent of spatial location.
From a mathematical perspective, this corresponds to a mean-field approximation in which the dispersal process effectively homogenizes recruitment across the domain. We therefore neglect spatial variations in the density and approximate the density field as spatially homogeneous,
$\rho(\mathbf{x},h,t)\equiv \rho(h,t)$.
Under this assumption, the recruitment boundary condition simplifies to
\begin{align}
\rho(h_0,t) = \kappa \int_{h_0}^{\infty} dh'\, (h')^{1+2H}\rho(h',t),
\label{eq:injection-rate-mf}
\end{align}
where
\begin{equation}
\kappa = \int d\mathbf{x}'\,\mathcal{K}(\mathbf{x}-\mathbf{x}')
\end{equation}
is the spatial integral of the dispersal kernel, which is independent of $\mathbf{x}$ under translational invariance.
Although the full time-dependent dynamics can be written as an implicit self-consistent solution \SInoref, the stationary regime admits an exact analytical solution, which we derive below.

\subsection*{Exact stationary solution reveals distinct dynamical regimes}
\begin{figure}
    \centering
    \includegraphics[
        alt={Prova3},
        width=\linewidth
    ]{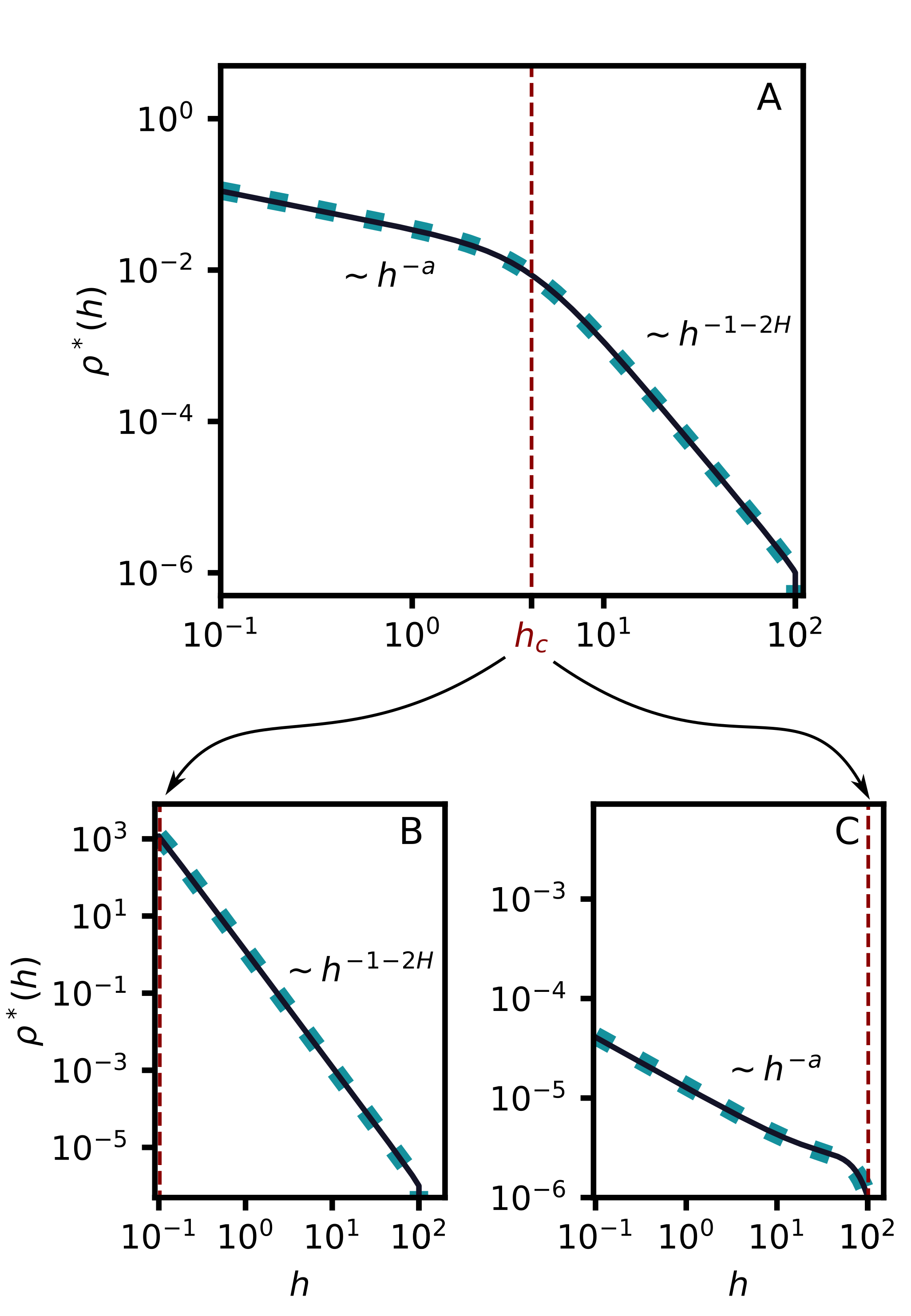}
    \caption{\textbf{Exact stationary solution for tree-size distribution shows the signature of the underlying dynamical regime and the local resource availability as a dual power-law shape.}
    \textbf{A}: In the spatial mean-field case, the stationary solution of \cref{eq:complete-growth-equation}, $\rho^*(h)$, as given in \cref{eq:statsol}, displays a dual power-law shape separated by a crossover height $h_c$. At lower tree sizes, the resource competition term dominates. In this size range, the distribution follows a power-law decrease with exponent $-a$, where $a=R^*/R_a$ is the resource consumption ratio. Above the crossover size $h_c$, the spatial competition term dominates and the power-law exponent becomes $-(1+2H)$, which is linked only to the tree allometric scaling relation. Parameters: $\kappa=0.001$ and $R_a=220$ (implying $\rho^*(h_0) \!\approx \! 0.11$, $a \! \approx \! 0.5$, and $h_c \! \approx \! 4.2$).
    \textbf{B}: If the available resources are very abundant and / or the sampling recruitment rate is high, the crossover point shifts towards very low values $h_c\to h_0$. In this scenario only the power law with exponent $-(1+2H)$ is visible. Parameters: $\kappa=10$ and $R_a=235$ ($\rho^*(h_0) \!\approx \! 1.2 \cdot 10^{3}$, $a \!\approx \!0.5$, and $h_c \! \approx \! 0.1$). \textbf{C}: If the available resources are scarce or seed injection / recruitment is low, the crossover point shifts to high values, $h_c\to h_u$. In this case, the only visible power law has exponent $-a$, with $a < 1+2H$. Parameters: $\kappa = 10^{-6}$ and $R_a=80$ (implying $\rho^*(h_0) \!\approx \! 4 \cdot 10^{-5}$, $a \!\approx \!0.5$, and $h_c\! \approx \! 101$). Common parameters across all panels: $H=1$, $h_0=h_b=0.1$, $\gamma=1$, $h_u=100$.}
    \label{fig:stationary}
\end{figure}
We now analyze the stationary solutions of the model, which characterize the long-term structure of the forest. To reduce the number of free parameters, we introduce the rescaled variables
\begin{equation}\label{eq:hatrho}
\hat\rho(h,\tau)=\frac{d}{g_0}\rho\!\left(h,\tau/g_0\right).
\end{equation}
For notational convenience, we continue to denote the rescaled density by $\rho$ rather than by $\hat\rho$, while the use of the rescaled time variable $\tau$ will indicate dimensionless quantities throughout the remainder of this section. Consequently, the parameters $\kappa$ and $\gamma$ are implicitly redefined in accordance with this rescaling. 
As we will show, the stationary tree-size distribution encodes the relative importance of growth, recruitment, and density-dependent competition for resources, canopy space, and light. We therefore seek time-independent solutions $\rho^*(h)$ of the rescaled form of \cref{eq:complete-growth-equation}.
This can be obtained in closed form in the general case \MMnoref. The full expression involves hypergeometric functions together with a self-consistent equation for the recruitment term. However, in the biologically relevant intermediate regime
$h_0 \ll h \ll h_u$, and for \(a \equiv R[\rho^*]/R_a\leq 1+2H\), the asymptotic expansion of the hypergeometric functions yields a considerably simpler and more interpretable expression for the stationary density:
\begin{align}
 \rho^*(h) \approx \left( \frac{h^a}{h_0^a \rho^*(h_0)} + (1+\gamma)\frac{h^{1+2H}}{1-a+2H} + \text{l.d.t.}\right)^{-1}, \label{eq:rhostar-general}
\end{align}
where $\rho^*(h_0)$ is computed self-consistently from the recruitment condition \cref{eq:injection-rate-mf} \MMnoref.

\Cref{eq:rhostar-general} consists of the sum of two competing power-law contributions, each associated with a distinct dynamical regime. As a consequence, the stationary tree-size distribution exhibits a crossover between two scaling behaviors at a characteristic height scale $h_c$.
By equating the two terms in \cref{eq:rhostar-general}, we obtain the crossover scale 
\begin{equation}
h_c = \left( \frac{1-a+2H}{(1+\gamma)\, h_0^a \,\rho^*(h_0)} \right)^{1/\left(1+2H-a\right)}.
\label{eq:crossover-term}
\end{equation}
As shown in \cref{fig:stationary}A, the stationary density therefore displays a double power-law behavior. In the regime $h_0 \ll h \ll h_c$, the first contribution dominates and $\rho^*(h)\sim h^{-a}$, whereas for $h_c \ll h \ll h_u$, the second term becomes dominant, yielding $\rho^*(h)\sim h^{-1-2H}$.
Therefore, the stationary size-distribution exhibits two distinct scaling regimes. For small individuals, $h<h_c$, the dynamics is primarily controlled by competition for shared resources, leading to the scaling $\rho^*(h)\sim h^{-a}$, with $a<1+2H$. As trees grow, spatial exclusion, light and canopy competition progressively become dominant, giving rise to the asymptotic scaling $\rho^*(h)\sim h^{-1-2H}$ for $h>h_c$.
This crossover is consistent with the well-established role of resource limitation during seedling establishment and juvenile growth, particularly under low-light understory conditions \cite{Tilman1982,PoorterKitajima2007,Zhu2025}.

The critical height $h_c$ depends on the the sapling recruitment rate and the local available resources $R_a$. For instance, if the sapling recruitment efficiency is high, then the injection term $\rho^*(h_0)$ will grow. In this case, for a given $R_a$, the crossover will shift toward lower values, eventually reaching the sapling height $h_c \to h_0$ \SInoref.  The same effect happens when $R_a\to\infty$. In both cases, the relative importance of the resource sink term decreases, and the stationary distribution displays a single power law with exponent $\approx -(1+2H)$ (\cref{fig:stationary}B). In the opposite limit of scarce resources and weak injection, the crossover point shifts to the right, eventually reaching $h_u$. When $h_c\to h_u$, then again $\rho^*(h)$ displays a single power law, this time with exponent $-a$ (\cref{fig:stationary}C). Interestingly, the influence of the canopy shading parameter $\gamma$ on the  crossover height is modest, compared to the other parameters \SInoref. 

We find that $a\leq (1+2H)$ \MMnoref, where $a\approx 1 + 2H$ indicates that the resource consumption is at its saturation value. In other words, if a single power law behavior is observed with (negative) exponent $a\ll1+2H$, then this indicates that the resource sink term is the limiting factor. Instead, if $a\approx1+2H$, then the forest growth is not limited by resource availability, but rather by space competition. 

Finally, regarding the stability of  the stationary solution discussed in this section, $\rho^*(h)$, taking the limit $\tau \to \infty$ in the formal time-dependent, spatially homogeneous exact solution to the mean-field \cref{eq:complete-growth-equation} for $\gamma=0$ \SInoref, indicates that this solution is globally attracting in the general case. Furthermore, numerical simulations show that the spatially homogeneous, translation-invariant system converges to the homogeneous solution $\rho^*(h)$ even when initialized from different random and non-random inhomogeneous conditions, supporting that it is globally stabile.

\subsection*{Effect of space and border disturbance}
We now relax the assumption of spatial homogeneity and reintroduce the explicit spatial dependence of the model. As discussed above, spatial coupling enters through the dispersal kernel $\mathcal K(\mathbf{x},\mathbf{x}')$, which describes recruitment at position $\mathbf{x}=(x_1,x_2)$ generated by individuals located at position $\mathbf{x}'$. We model dispersal using the simplest isotropic kernel with a characteristic length scale, namely an exponential kernel with correlation length $\xi$, and consider a forest distributed on a homogeneous two-dimensional surface. In the absence of explicit boundaries, the spatially homogeneous steady state $\rho^*(\mathbf{x},h)=\rho^*(h)$ remains a stationary solution of the spatially explicit model \MMnoref.

\begin{figure*}
    \centering
    \includegraphics[
        alt={Prova3},
        width=\linewidth
    ]{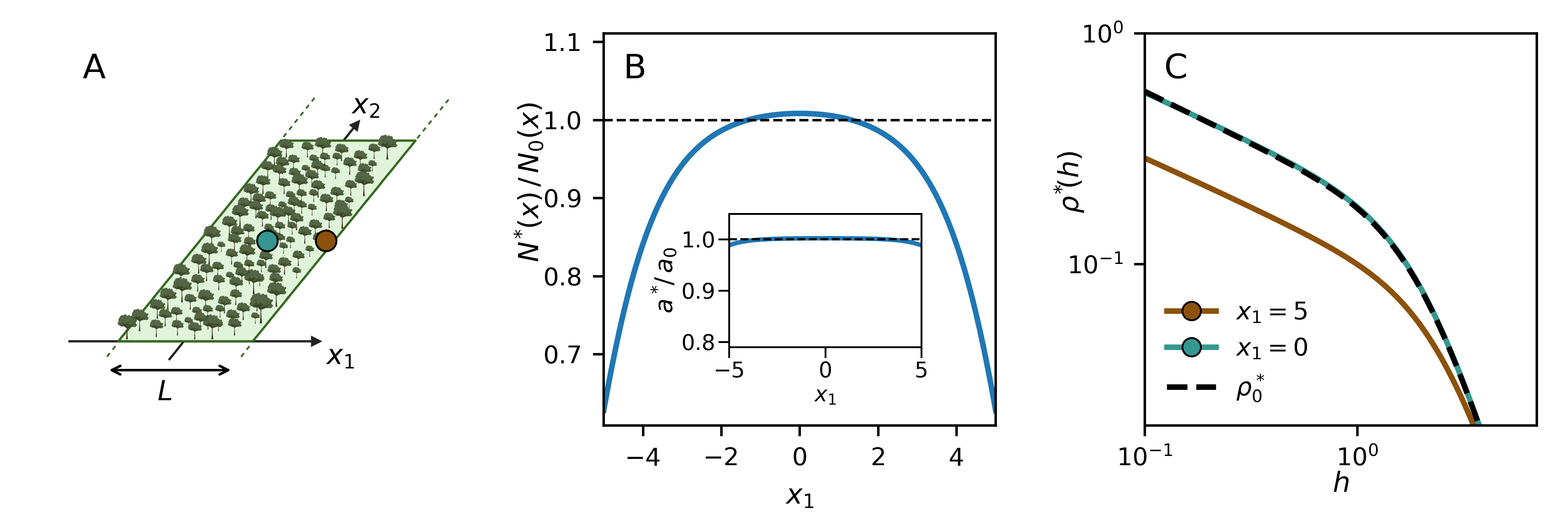}
    \caption{\textbf{Spatial coupling causes border disturbance, which disproportionately affects smaller trees.}
    \textbf{A}: Depiction of border disturbance: starting from the spatially homogeneous solution, the forest is then confined to a land strip, as in \cref{eq:forest-strip}. The two colored markers mark bulk and edge positions for which the stationary density is shown in panel C.
    \textbf{B}: Stationary post-disturbance total tree density $N^*(x)= \int dh\,  \rho^*(x,h)$ relative to pre-disturbance state $N_0 =  \int dh\,\rho_0^*(h)$. Inset: stationary scaling exponent $a^*(x)$ relative to pre-disturbance exponent $a_0$ as a function of the spatial coordinate $x_1$.
    \textbf{C}: Stationary post-disturbance distribution in the bulk (green line) and border (brown line) compared to initial pre-disturbance level (black dashed line). Parameters: $H=1$, $h_0=h_b=0.1$, $\kappa=0.01$, $R_a=125$, $h_u=50$, $\xi=1$, and $L=10$.}
    \label{fig:border}
\end{figure*}

Starting from this homogeneous steady state, we investigate the effect of a boundary disturbance in a particularly simple geometry. At $t=0$, we impose
\begin{equation}
\rho(\mathbf{x},h,t)=0 \qquad \text{for} \qquad |x_1|>\frac{L}{2}\quad\text{and}\quad t\geq0,
\label{eq:forest-strip}
\end{equation}
thereby representing a region in which the forest has been completely removed by an external disturbance and regrowth is prevented. Equivalently, the forest is constrained to grow within an infinite strip of width $L$, aligned along the $x_2$ direction (\cref{fig:border}A).
By translational invariance along the $x_2$ direction, we can integrate one spatial dimension and reduce the problem to a one-dimensional system with a different dispersal kernel (a Bessel function, see \MM).

\Cref{fig:border}B shows the relative variation in the total tree density with respect to the homogeneous stationary state.
Near the boundary, the total tree density is strongly reduced due to the suppression of seed influx from the removed side of the forest.
Interestingly, although the total tree density decreases by almost $40\%$ near the boundary, the relative variation in the resource saturation parameter, ${a^*}/{a_0}$,
remains only a few percent  (\cref{fig:border}B, inset). The reason for this behavior is shown in \cref{fig:border}C: the dominant contribution to the depletion of the total density originates from small individuals, as demonstrated by the comparison between the size distributions in the bulk and near the boundary. This behavior is consistent with the fact that smaller trees are more strongly controlled by the recruitment term and are therefore more sensitive to reductions in seed injection. By contrast, the large-tree sector of the distribution, corresponding to the regime above the crossover scale $h_c$, is only weakly affected by the boundary perturbation. As a consequence, the total resource consumption remains nearly uniform across the system.  This reflects the fact that the resource-consumption term is dominated by larger trees, whose distribution is comparatively insensitive to edge effects. 

As a final remark, we note that the effect of the boundary has a range consistent with the spatial scale of the kernel, and is mostly disappeared at a distance of $\approx 3$ from the boundary (space is measured in units of $\xi=1$). We now show that this picture can change drastically in the presence of different species with different dispersal lengths.

\subsection*{Interplay of boundary effects and inter-species interactions}
To reintroduce the additional complexity associated with multiple tree species avoiding proliferation of parameters, we consider the case of two species (or species groups) characterized by different dispersal properties but identical growth and mortality dynamics.
We retain the exponential kernel introduced in \cref{eq:exp-kernel}, while allowing for species-dependent dispersal lengths $\xi_\alpha$, with $\alpha=1,2$. Importantly, the normalization of the kernel ensures that its spatial integral is independent of $\xi_\alpha$. As a consequence, in a homogeneous two-dimensional environment, the spatially homogeneous stationary solution $\lambda_\alpha \rho_\alpha^*(x, h)$, with $\sum_\alpha\lambda_\alpha=1$
remains a stationary solution for each species, independently of their dispersal range \MMnoref.
This property reflects the fact that, in the absence of spatial heterogeneity or boundaries, the total recruitment effort remains balanced across species. Ecologically, this is consistent with the well-known trade-offs associated with seed dispersal strategies, whereby broader dispersal spreads recruitment over larger spatial scales while reducing local recruitment density, and physical properties of heavier seeds that tend to travel shorter distances \cite{Nathan2000,Chen2020,Treep2021}.

\begin{figure*}
\centering
\includegraphics[width=\linewidth]{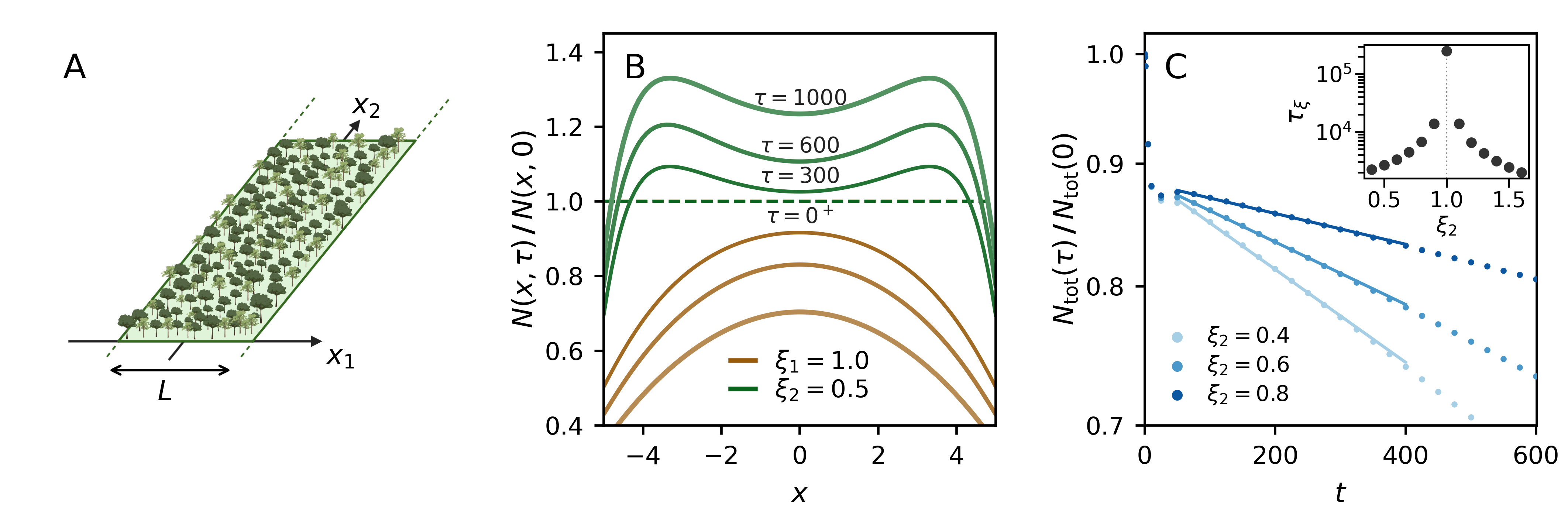}
\caption{\textbf{Interplay between dispersal and inter-species competition amplifies border disturbance inducing competitive exclusion between species with different dispersal ranges.}
\textbf{A}: Depiction of border disturbance: same spatial configuration as in \cref{fig:border}, but with two competing species with different dispersal ranges $\xi_1=1, \xi_2$.
\textbf{B}: Spatial profile of the total tree density per unit area $N_\alpha(x)$ for both species, shown at different times after the introduction of a boundary disturbance. While both species initially occupy the strip homogeneously (dashed line), the species with the larger dispersal range (brown solid lines) progressively decreases near the boundary and eventually throughout the entire system, whereas the more localized species becomes dominant (green solid lines).
\textbf{C}: Total tree density of the disadvantaged species with fixed dispersal length $\xi_1=1$ as a function of time, for different values of $\xi_2<\xi_1$. The abundance of the disadvantaged species decreases approximately exponentially, with a decay rate controlled by the mismatch between the two dispersal scales.
Inset: Characteristic decay timescale $\tau_\xi$ of the disadvantaged species as a function of the dispersal length ratio, computed from an exponential fit (note the log y-scale) in the range $50\leq \tau \leq 300$. The timescale diverges as $\xi_2\to\xi_1$, where the two species become dynamically equivalent (vertical dotted line). Spatial distances are measured in units of the reference dispersal scale $\xi_1=1$. Parameters: $H=1$, $h_0=h_b=0.1$, $\kappa=0.018$, $R_a=112$, $h_u=50$, $\xi=1$, and $L=10$.}
\label{fig:fig5}
\end{figure*}

We consider the same spatial configuration introduced in the previous subsection and defined in \cref{eq:forest-strip}, namely a forest strip of width $L$ surrounded by uncolonizable terrain, starting from the homogeneous stationary distribution (\cref{fig:fig5}A). For symmetry, we choose equal starting densities for the two species.
\Cref{fig:fig5}B shows the spatial profile of $N_\alpha(x, \tau)=\int \mathrm dh\rho(x,h,\tau)$, the total tree density per unit area for the two species, at different times, normalized by the initial density. Immediately after initialization, the two profiles are nearly identical and spatially uniform, consistently with the homogeneous stationary state. As time evolves, however, the species characterized by the longer dispersal length progressively decreases near the boundary, while the more localized species increases its density.
Interestingly, this trend persists over very long timescales, with the long-dispersal species continuing to decline in relative abundance throughout the system's evolution. Moreover, the spatial extent of this effect significantly exceeds the characteristic dispersal length: while the dominant species displays a boundary layer comparable to the single-species case shown in \cref{fig:border}, the density reduction of the declining species extends across the entire strip.
\Cref{fig:fig5}C shows $N_{\rm tot}(\tau)=\int\mathrm d x N_1(x,\tau)$, the total tree density per unit area of the species with $\xi_1=1$,  as a function of time for different values of $\xi_2<\xi_1$. In all cases, the density undergoes an initial decay that is approximately exponential, with a rate that depends on the mismatch between the two dispersal scales, before eventually settling at very low values \SInoref.
Finally, the inset in \cref{fig:fig5}C shows the characteristic decay timescale $\tau_\xi$, defined as the inverse decay rate of the declining species (species 1 for $\xi_2<\xi_1=1$, and species 2 otherwise), as a function of $\xi_2$.
The decay timescale increases significantly as $\xi_2$ approaches $\xi_1$ from both sides, and then diverges as $\xi_2\to\xi_1$, reflecting the fact that the two species become dynamically equivalent in this limit.

Taken together, the results shown in \cref{fig:fig5} indicate that species with broader dispersal kernels are more sensitive to boundary effects, leading to a substantial reduction in their total abundance. This behavior originates from the suppression of recruitment near the boundaries, which primarily affects smaller individuals and therefore disproportionately impacts the species relying more strongly on long-range dispersal. The resulting depletion of juvenile trees weakens the competitive balance of the long-dispersal species, ultimately favoring the expansion and dominance of the more localized competitor.

\section*{Discussion}
In this work, we introduced a spatially explicit, size-structured model for forest dynamics that combines ontogenetic growth, canopy shading effects, density-dependent competition, and non-local recruitment through seed dispersal. 
A central feature of our framework is that, starting from metabolic scaling principles, it links three levels of description that are often treated separately: individual metabolic growth, stand-level size structure, and spatially non-local recruitment. Our approach remains sufficiently mechanistic while retaining analytical tractability. As a result, the scaling regimes emerging in the tree-size distribution arise directly from explicit demographic processes rather than from phenomenological assumptions, allowing observed forest structure to be interpreted in terms of underlying ecological mechanisms.
 
In a spatially homogeneous setting, the model has an exact solution for the stationary tree-size distribution, characterized by a crossover between two distinct power-law regimes. For small individuals, the distribution is controlled primarily by resource competition and depends explicitly on the degree of resource saturation. For larger trees, the dynamics becomes dominated by spatial shading and canopy competition, leading to a scaling exponent determined solely by crown allometry. The crossover scale separating these regimes therefore provides a direct connection between measurable forest structure and the dominant processes regulating population dynamics. In particular, the emergence of a single power-law regime may indicate whether forest dynamics are primarily constrained by resource limitation or by space occupation, suggesting that tree-size spectra may contain information about the underlying ecological state of the system.
From an ecological perspective, these results are consistent with resource limitation dominating juvenile stages \cite{Kobe1995, Nathan2000,PoorterKitajima2007,Chen2020,Treep2021} and spatial competition becoming increasingly important for larger individuals. Importantly, this connection between the dynamical regimes and a measurable quantity, the tree-size scaling exponent, provides a possible way of a qualitative estimate of a forest condition, and is in qualitative agreement with empirical studies linking the exponent of tree-size distributions to the degree of disturbance \cite{Anfodillo_2013,Sellan_Simini_2017,Eichenwald_2025}. Within our framework, these variations acquire a mechanistic interpretation, reflecting shifts in the relative importance of resource limitation and spatial competition.

By reintroducing explicit spatial dependence, we further showed that finite boundaries induce substantial deviations from the homogeneous stationary state. In the single-species case, finite boundaries suppress local recruitment and primarily affect the density of small individuals, while leaving the large-tree sector comparatively unchanged. As a consequence, substantial reductions in total tree abundance can occur even when overall resource consumption remains approximately constant. This result highlights the distinct ecological roles played by different size classes and suggests that demographic responses to fragmentation or edge effects may be concentrated disproportionately among younger individuals, possibly providing a new theoretical tool to interpret field reports of habitat fragmentation and local disturbances \cite{Laurance1998Recruitment,Ewers2006,NunezAvila2013}. 
    
The multi-species model reveals an additional mechanism emerging from the interaction between dispersal and competition.  Dispersal-related life-history trade-offs have long been recognized as important determinants of coexistence in heterogeneous landscapes \cite{D_Andrea_JODwyer_2021}. Here we show that, even in homogeneous environments, explicit spatial boundaries can break the resulting symmetry between dispersal strategies and drive long-term competitive imbalance.
Species characterized by broader dispersal kernels experience a stronger reduction in recruitment near boundaries, progressively weakening their competitive position and favoring species with shorter dispersal ranges. Interestingly, this effect extends well beyond the intrinsic dispersal scale, producing system-wide shifts in abundance despite originating from a localized perturbation. At the same time, these results should not be interpreted as implying inevitable exclusion of long-dispersal species. When dispersal differences are small, the characteristic exclusion timescale becomes extremely long, reaching hundreds of multiples of the individual growth timescale $g_\alpha^{-1}$. Such time-scales would be much longer than those over which the environmental conditions stay stationary, and extend beyond the expected range of validity of our framework.
Moreover, real forests are characterized by multiple ecological trade-offs involving growth rate, shade tolerance, fecundity, resource-use efficiency, and disturbance resilience \cite{Westoby1996,MullerLandau2010,Diaz2015}. Such trade-offs may counterbalance the boundary-mediated disadvantages identified here and generate coexistence, successional dynamics, or spatial niche partitioning \cite{Levine2009,Wright2010,Adler2014,Jops2023}.

Previous theoretical studies have successfully combined size structure and demographic stochasticity within neutral frameworks to investigate emergent patterns of community organization \cite{o2009integrative}. More generally, scaling approaches have shown that universal tree-size distributions can emerge from allometric and energetic constraints, including through optimization principles \cite{simini2010self}. The present framework provides a complementary dynamical perspective by explicitly incorporating ecological interactions. Within this formulation, the universal scaling exponent is recovered as the limiting case in which forest dynamics are no longer constrained by resource availability, while departures from this exponent arise naturally from resource limitation. This establishes a mechanistic connection between the observed scaling exponent and the underlying dynamical regime of the forest, providing conditions under which the optimal scaling state can be attained. More broadly, our framework allows forest size distributions and spatial abundance patterns to be interpreted as signatures of the ecological processes shaping forest dynamics.

The present study should nevertheless be regarded as a first step aimed at isolating the basic dynamical consequences of coupling growth, competition, and dispersal within a tractable framework. Several biologically relevant ingredients have intentionally been neglected. In particular, we assumed identical growth and mortality dynamics across species, homogeneous environmental conditions, and isotropic dispersal kernels. Resource limitation was represented through an effective collective variable, and did not explicitly distinguish among different limiting resources such as water or nutrients \cite{Bloom_1985,Chapin_1987,PoorterKitajima2007,Umarani_2024}. Environmental heterogeneity \cite{chesson2000mechanisms,Melbourne_2008,Padmanabha2024}, demographic fluctuations \cite{o2009integrative,Lande_1993,Bowler2022}, landscape geometry and fragmentation \cite{Laurance_2002,HARPER_2005,Borthagaray_Arim_Marquet2012,BernardiDoimo2026}, and temporally varying resource availability \cite{Wright_2002,Adler_2006,Levine2009,Jops2025} are all expected to influence forest dynamics and may modify the scaling regimes identified here as well as the strength of boundary-mediated competitive effects. Incorporating these ingredients while preserving analytical tractability represents an important direction for future work and may provide a bridge between mechanistic forest theory and empirical observations of forest structure, biodiversity, and ecosystem resilience.

\section*{Materials and Methods}

\subsection*{Growth rate} We derive the growth term $g_{\alpha}(h)$ introduced in \cref{eq:single-tree-growth}. 
Following established metabolic scaling theory, the temporal dynamics of the metabolically active biomass $M$ of an individual organism can be described via the energy balance equation \cite{west1997general, enquist1998allometric, enquist2002global, brown2004toward, west2001general, mori2010mixed}:
\begin{equation}
\frac{d M}{dt} = c^{\alpha}_{1} B - c^{\alpha}_{2} M,
\label{eq:single-mass-eq}
\end{equation}
where $B$ is the total metabolic rate, and $c^{\alpha}_1, c^{\alpha}_2$ are species-specific coefficients. We assume that the assimilation coefficient $c_1^{\alpha} = \epsilon_1 R_{a} / (\zeta + R_{a})$ depends on the available resources $R_{a}$, where $\zeta$ is a half-saturation constant, whereas the maintenance cost coefficient is scaled as $c^{\alpha}_2 = \epsilon_2 / h^{\alpha}_{M}$, with $h^{\alpha}_{M}$ denoting the theoretical maximum height achievable by species $\alpha$.
Supported by empirical and theoretical evidence \cite{Comita2007, simini2010self}, we model the allometric scaling between the crown radius $r_c$ and tree height $h$ as $r_c \sim h^H$, where $0 \leq H \leq 1$ is the Hurst exponent. Consequently, the crown volume scales as $V \sim h \times r_c^2 \sim h^{2H +1}$. As detailed in the Main Text, the metabolic rate is proportional to the crown volume, yielding $B \sim h^{2H +1}$. Assuming that the sap flux velocity at the trunk basis is approximately independent of tree size \cite{wullschleger2000radial}, the conducting cross-sectional area $\Sigma_c$ must scale proportionally to the metabolic rate ($\Sigma_c \sim B$). The active biomass is then proportional to the product of the conducting area and the tree height, leading to $M \sim \Sigma_c \times h \sim B h \sim h^{2H+2}$ \cite{simini2010self}.
Substituting these scaling relations into the energy balance equation \cref{eq:single-mass-eq} and expressing the active mass $M$ as a function of height $h$, we obtain
\begin{equation}
\frac{dh_{\alpha}(t)}{d t}=g_\alpha(h_\alpha(t))\equiv g_0^{\alpha}\left(1-\frac{h_\alpha(t)}{h_u^{\alpha}}\right),
\end{equation}
where $g_0^{\alpha}= c^{\alpha}_1/(2H+2)$ and $h_u^{\alpha}=c^{\alpha}_1/c^{\alpha}_2$ are species-specific parameters. Note that the growth term $g_{\alpha}(h)$ does not depend explicitly on time. The asymptotic height $h^{\alpha}_u$ is given by the following:
\begin{equation}
    h^{\alpha}_u = \frac{R_{a}}{\zeta + R_{a}} \frac{h^{\alpha}_{M}}{\epsilon^{\alpha}_2},
    \label{eq:hu-vs-hmax}
\end{equation}
implying that in the case of poor available resources $R_{a}$, the plant stops growing before reaching the maximum height $h^{\alpha}_{M}$, consistent with empirical observations. 

\subsection*{Light attenuation and shading}
Taller individuals intercept part of the incoming radiation, reducing the amount of light that reaches shorter plants. We account for this effect by assuming that the growth rate is modulated by the local effective light availability $I_{\alpha}(\mathbf{x},h,t)$:
\begin{equation}
    \tilde{g}_{\alpha}(\mathbf{x},h,t) = I_{\alpha}(\mathbf{x},h,t)\,g_{0}^{\alpha} \left(1-\frac{h}{h_u}\right)
\end{equation}
where plants taller than $h$ contribute to shading the plants below them. Assuming that light attenuation follows the Beer-Lambert law \cite{monsi1953lichtfaktor,MONSI2005}, $I_{\alpha}(\mathbf{x},h,t)$ decays exponentially with the amount of vegetation crossed by the incoming light:
\begin{equation}
I_{\alpha}(\mathbf{x},h,t)=I_0\exp\left\{-\gamma_{\alpha}
\int_h^{\infty}\rho(\mathbf{x},h',t)\,h'^{2H}\,dh'
\right\},
\end{equation}
where $\gamma_{\alpha}$ is the species-specific light attenuation coefficient, capturing both the physical shading effect of standing biomass and how efficiently species $\alpha$ exploits the available light, and the integral accounts for the cumulative shading exerted by all plants taller than $h$.

\subsection*{Canopy interaction} The mortality of this term is due to direct interactions between tree canopies, and their effects on the evapotranspiration rate. Because this term is due to spatial interference of tree canopies, the interaction range of this term will be proportional to the lateral size of the crown, $h^H$. Consequently, the mortality experienced by a tree of height $h$ is proportional to the local density of individuals of comparable size within this interaction area. Specifically, for a plant of height $h$, the space competition $d_S$ at position $x$ and time $t$ is 
\begin{equation}
    d^{S}_{\alpha}(\mathbf{x},h,t) = d_\alpha I_\alpha(\mathbf{x}, h, t) \cdot h^{2H} \rho(\mathbf{x},h,t),
\end{equation}
where $\rho(\mathbf{x},h,t)=\sum_{\alpha}\rho_{\alpha}(\mathbf{x},h,t)$. Since the evapotranspiration rate depends on the available light intensity, we take this term as proportional to it.

\subsection*{Resource competition} In the limit of negligible spatial competition, $d^{S}=0$, the resources consumed by the forest per unit time obey the following equation
\begin{equation}
\begin{split}
\dot{R}\left[\rho_\alpha(\mathbf{x},t)\right] &= \left(1-\frac{R\left[\rho_\alpha(\mathbf{x},t)\right]}{\tilde{R}_a}\right)(1+2H) \\
    &\quad\times \int_{h_0}^{\infty} dh\, h^{2H}\, \tilde{g}_\alpha(h)\, \rho_\alpha(\mathbf{x},h,t) \\
    &\quad + B_\alpha(\mathbf{x},t),
\end{split}
\label{eq:time-evolution-resources}
\end{equation}
where 
\begin{equation}
\begin{split}
    B_{\alpha}(\mathbf{x},t) &\equiv -h^{2H+1}\, \tilde{g}_{\alpha}(h)\, \rho_{\alpha}(\mathbf{x},h,t)\Bigg\vert_{h_0}^\infty \\&= h_0^{2H+1}\, \tilde{g}_{\alpha}(h_0)\, \rho_{\alpha}(\mathbf{x}, h_0,t)\geq 0,
\end{split}
\label{}
\end{equation}
where we have taken into account that $\rho_\alpha(\mathbf{x},h,t)=0$ when $h>h_u$.
\Cref{eq:time-evolution-resources} implies that, neglecting the boundary contribution $B_\alpha$, the total resource consumption $R[\rho_\alpha(\mathbf{x},t)]$ increases whenever $R[\rho_\alpha(\mathbf{x},t)]<\tilde{R}_a(\mathbf{x})$ and decreases whenever $R[\rho_\alpha(\mathbf{x},t)]>\tilde{R}_a(\mathbf{x})$. Therefore, the dynamics tends to drive the system toward a state in which resource consumption balances the available resource flux. The boundary term $B_\alpha$ becomes relevant only in the presence of very strong recruitment, in which case it merely shifts the equilibrium value of the resource consumption.
For simplicity, in the main text we redefine $R_a = \tilde{R}_a/(1+2H)$.

\subsection*{Stationary solution}
We derive the stationary solution of the model in the spatially homogeneous case, under the neutral assumption of identical growth and resource-consumption parameters across species. Specifically, we look for $\rho^*(h)$, a time-independent solution of the following PDE for $\rho \equiv \rho(h,\tau)$ (see \cref{eq:hatrho}):
\begin{widetext}
\begin{align}
   \! \!\partial_\tau&\rho =-\!I\!\left\{\!\partial_h\! \left[\!\left(1\!-\!\frac{h}{h_u}\!\right)\rho\right] \!+\!\left[\!\gamma \left(\!1\!-\!\frac{h}{h_u}\!\right)\!+\!1\!\right]h^{2H}\!\rho^2 \!+\! \left(1\!-\!\frac{h}{h_u}\!\right)\!\frac{a(\tau)}{h}\rho\!\right\},
    \label{eq:simplified-growth-equation}
\end{align}
\end{widetext}
where, for notational convenience, we have redefined the parameter $\gamma g_0/d$ as $\gamma$ and 
\begin{equation}\label{eq:a-2}
    a(\tau)\equiv \frac{R[\rho(\tau)]}{R_a},\quad R[\rho(\tau)]\!\equiv\! \int_{h_0}^\infty \!\!\!\!h'^{1+2H}\rho(h',\tau)\,dh'
\end{equation} 
and $ d/g_0 \,R_a$ as $R_a$.
The stationary solution of \cref{eq:simplified-growth-equation} can be obtained by a suitable change of variable \SInoref, which leads to the following solution
\begin{widetext}
\begin{equation}\label{eq:statsol}
\begin{split}
    \! \frac{1}{ \rho^*(h)}\!  =\!  \left(\! \frac{h}{h_0} \right)^a\!  \Bigg[\frac{1\! -\! h/h_u}{1\! -\! h_0/h_u} \frac{1}{ \rho^*(h_0)} +\!h_0^a (1-h/h_u)\! \int_{h_0}^{h} dh' \! h'^{2H-a}\frac{1+\gamma(1-h'/h_u)}{\left(1-h'/h_u \right)^2}\! \Bigg],
    \end{split}
\end{equation}
\end{widetext}
where $a$ denotes the steady-state resource consumption ratio, as determined from \cref{eq:a-2} evaluated at $\rho=\rho^{*}$, and $\rho^*(h_0)$ is computed self-consistently from the recruitment condition \cref{eq:injection-rate-mf}. The last equation can be  re-expressed in terms of hypergeometric functions ${}_2F_1$
\begin{align}
\int_{x_0}^{x} dz\,
z^{\delta-1}\frac{1+\gamma-\gamma z}{(1-z)^2} = F(x)-F(x_0),
\end{align}
where 
\begin{align}
    F(z)= \frac{z^{\delta}}{\delta} \left[
{}_2F_1\!\left(2,\delta \delta+1;z\right) + \gamma\,{}_2F_1\!\left(1,\delta;\delta+1;z\right) \right]
\end{align}
and $\delta = 1-a+2H$.
In this limit,  if $a<2H+1$, the hypergeometric function simplifies to its leading-order power law,
so that the stationary density, is given by \cref{eq:rhostar-general}, and is governed by the competition between two distinct power-law scaling regimes, as explained in the main text.

\subsection*{Homogeneous solutions of the spatially explicit model}
We consider a homogeneous two-dimensional environment in which species differ only through their dispersal kernels. We show that, in the absence of explicit boundaries or environmental heterogeneity, the spatially explicit model admits homogeneous stationary solutions that are independent of the dispersal range.
Let $\rho_\alpha^*(\mathbf{x},h)=\rho_\alpha^*(h)$
be a stationary density independent of position. The stationary equations become
\begin{widetext}
\begin{align}
&\!\!\!0 =\!-\partial_h\!\left[g(h)\rho_\alpha^*(h)\right]\!-\!\left[\left(\gamma g(h)\!+\!1\right)h^{2H}\!\rho^*(h)\!+\!\frac{g(h)}{h}a^*\right]\rho_\alpha^*(h), \\
&\!\!\!\rho_\alpha^*(h_0) = \int d\mathbf{x}'\mathcal{K}_\alpha(\mathbf{x}-\mathbf{x}') \int_{h_0}^{\infty}\!\!dh'\,h'^{1+2H}\rho_\alpha^*(h').
\end{align}
\end{widetext}
Here, we define $\rho^*(h) = \sum_\beta \rho_\beta^*(h)$, $g(h)=1-h/h_u$ and the quantity $a^*$ is specified by \cref{eq:a-2}, with $\rho(h',\tau)$ replaced by $\rho^*(h)$.
Since the dispersal kernel is normalized, $\int d\mathbf{x}'\,\mathcal{K}_\alpha(\mathbf{x}-\mathbf{x}')=\kappa,$
independently of the dispersal length. Therefore, for spatially homogeneous densities, the recruitment term depends only on the total reproductive output of the species and not on the shape of its dispersal kernel.
Summing the stationary equations over all species yields
\begin{align}
&0 =\!-\partial_h\!\left[g(h)\rho^*(h)\right]\!-\!\left(\gamma g(h)\!+\!1\right)h^{2H}{\rho^*}^{2}(h)\!-\!\frac{g(h)}{h}a^*\rho^*(h),\\
&\rho^*(h_0) = \kappa \int_{h_0}^{\infty}\!\!dh'\,h'^{1+2H}\rho^*(h').
\end{align}
which coincides with the single-species stationary problem discussed in the previous section.
This implies that the per-species stationary density taken as proportional to the total density,
\begin{equation}
    \rho^*_{\alpha}(h) = \lambda_{\alpha}\,\rho^*(h),
    \qquad \lambda_\alpha \geq 0,\quad \sum_{\alpha} \lambda_{\alpha}=1,
\end{equation}
is itself a stationary solution. Thus, in a homogeneous environment, species characterized by different dispersal ranges remain dynamically equivalent, and the stationary state is determined only by their relative abundances $\lambda_\alpha$.

\subsection*{Spatial kernel used in simulations}
In the simulations shown in \cref{fig:border,fig:fig5}, we model dispersal using an exponential kernel with correlation length $\xi$,
\begin{equation}
\mathcal K_\xi\!\left(|\mathbf{x}-\mathbf{x}'|\right) = \frac{\kappa}{2\pi \xi^2} \exp\left(-\frac{|\mathbf{x}-\mathbf{x}'|}{\xi} \right).
\label{eq:exp-kernel}
\end{equation}
We consider an environment (the forest strip) which is translationally invariant along the $x_2$ direction. The problem therefore reduces effectively to a one-dimensional system with dispersal kernel
\begin{align}
\begin{split}
G\!\left(|x_1-x_1'|\right) &= \int dx_2 \int dx_2'\, \mathcal K(\mathbf{x},\mathbf{x}') \\
&= \frac{\kappa|x_1-x_1'|}{\pi\xi^2} K_1\!\left(\frac{|x_1-x_1'|}{\xi}\right),
\end{split}
\end{align}
where $K_1$ is the modified Bessel function of the second kind of order one.

\bibliography{refs}

\end{document}